\documentclass[aps,prl,groupedaddress,showpacs,showkeys]{revtex4}

\usepackage{pstcol,pst-node,pst-coil,pst-text,graphicx}

\usepackage{amsmath, amssymb, amsthm, alltt, setspace,bbm}

\def\ket#1{\left\vert #1 \right\rangle}
\def\bra#1{\left\langle #1 \right\vert}

\begin{document}


\title{From Dirac to Diffusion: Decoherence in Quantum Lattice Gases}

\author{Peter J. Love\footnote{Current Affiliation: D-Wave Systems Inc., Suite 320
1985 West Broadway, Vancouver, British Columbia, V6J 4Y3 Canada\\ Email: peter@dwavesys.com}}
\author{Bruce Boghosian}
\affiliation{Department of Mathematics, Tufts University, Medford, Massachusetts 02155 USA}

\date{\today}

\begin{abstract}
We describe a model for the interaction of the internal (spin) degree of freedom of a quantum lattice-gas particle with an environmental bath. We impose the constraints that the particle-bath interaction be fixed, while the state of the bath is random, and that the effect of the particle-bath interaction be parity invariant.  
The condition of parity invariance defines a subgroup of the unitary group of actions on the spin degree of freedom and the bath. We derive a general constraint on the Lie algebra of the unitary group which defines this subgroup, and hence guarantees parity invariance of the particle-bath interaction. We show that generalizing the quantum lattice gas in this way produces a model having both classical and quantum discrete random walks as different limits. We present preliminary simulation results illustrating the intermediate behavior in the presence of weak quantum noise.
\end{abstract}


\maketitle

\section{Introduction}

Lattice gases are arguably the simplest models for the simulation of classical physical systems. These models provide elementary microscopic dynamics whose hydrodynamic limits are, {\em inter alia}, the diffusion equation, Burgers' equation and the Navier Stokes equations~\cite{bib:doolenreview}. They also provide a simple arena for the creation of new models of physical phenomena such as multicomponent flow and dynamical geometry~\cite{bib:bcp,bib:lovepart,bib:lovedg,bib:meyer7}. Lattice gases possess deterministic, stochastic and quantum (unitary) formulations.

Simulation of quantum systems on quantum computers remains one of the very few applications for which exponential speedup over classical computation is provable. The quantum lattice gases defined by Meyer~\cite{bib:meyer8} and by Boghosian and Taylor~\cite{bib:bogwash} may be simulated on a quantum computer in exponentially fewer steps than are required on a classical computer and with an exponential reduction in hardware. These models, together with other approaches by Lloyd~\cite{bib:lloyd2,lloyd3} and by Ortiz and Gubernatis~\cite{bib:gubernatis1,bib:gubernatis2}, have made concrete Feynman's observation that simulation of quantum systems is hard on classical computers, but easy on quantum computers~\cite{bib:feynman1,bib:feynman2,bib:feynman3}.

Meyer was led to the definition of the quantum lattice gas by consideration of quantum generalizations of cellular automata (CA). The simplest possible quantum CA model would be a map in which the updated state of a cell depended linearly on the state of its neighbours and the global evolution rule was unitary. Meyer showed that the only such maps are trivial, namely, the identity map and the left or right shift, possibly multiplied by a phase.  This ``No-Go'' theorem proves that there are no nontrivial homogeneous scalar unitary cellular automata~\cite{bib:meyer1}. 

Quantum cellular automata models may evade this No-Go theorem~\cite{bib:meyer1} by having cell values which are not scalar, or by having local update rules which are not linear (although the global evolution in such models remains quantum  mechanical, and therefore linear). Non-scalar models sacrifice some simplicity, while the determination of unitarity for nonlinear models is problematic~\cite{bib:durr1,bib:durr2,bib:meyer9}. A third possibility exists: We may {\em partition} our cellular automata, dividing our evolution into two substeps, acting on two distinct neighborhoods. This was the approach taken by Meyer~\cite{bib:meyer2} and also by Watrous~\cite{bib:watrous}. If one of the substeps of the evolution is interpreted as propagation of the cell values to neighboring sites we may identify the partitioned cellular automata with a lattice gas model~\cite{bib:meyer2}.

The dynamics of all lattice gases take place by propagation of particles to neighboring sites on the lattice, followed by  a local collision operation. In a stochastic lattice gas which obeys semi-detailed balance, the collision step is a doubly-stochastic Markov matrix acting on the state of a single lattice site. For a lattice gas with only a single particle moving on the lattice, the stochastic model reduces to a classical random walk. In the collision step of a quantum lattice-gas model, the state at a site is acted on by a unitary scattering matrix. Just as in the classical case, the one-particle sector of the model possesses an interpretation as a discrete-time discrete-space quantum random walk.

Quantum lattice gases are explicitly formulated as discrete models for quantum physics, and so are subject to additional physical constraints.  In particular, the unitary scattering matrix is constrained to be parity invariant. The quantum lattice gas was shown to yield the continuum propagator for the Dirac equation in one dimension~\cite{bib:meyer2}, and also possesses the Schroedinger equation as a nonrelativistic continuum limit~\cite{bib:bogwash3}.

The quantum random walk has also been studied extensively from a point of view quite different from that of physical modeling. Quantum random walks may provide an alternative route for the development of new quantum algorithms.  Such walks have the property that the variance of the walk grows linearly with time, in sharp contrast to classical walks where the variance grows as the square root of time. This echoes the computational advantage of Grover's search algorithm, and indeed unstructured search can be reformulated as a quantum random walk problem.  Discrete-time quantum walks have yielded exponential performance improvements in the hitting time on the hypercube, and continuous-time quantum walks have yielded an exponential improvement in the solution of a graph traversal problem. For an overview of the subject of quantum random walks we refer the reader to the review of Kempe~\cite{bib:qrwreview}.

Two things distinguish quantum lattice gases from discrete quantum walks. Firstly, the absence of scalar homogeneous models means that discrete quantum walks must also introduce an extra degree of freedom. This degree of freedom is interpreted as a ``coin'' state which determines the motion of the walker at the next time step. In the context of the quantum lattice gas this degree of freedom is interpreted as the spin (or, in one dimension, helicity) degree of freedom of the particle. Secondly, the constraint of parity invariance imposed by Meyer in one dimension, and discrete rotation invariance imposed by Boghosian and Taylor in $d$-dimensions is infrequently applied to discrete time quantum random walks. The coin state is commonly updated by the Hadamard operation, which is not invariant under parity inversion. While parity inversion symmetry is a fundamental requirement for models of physics, from the point of view of the computational properties of quantum walks parity invariance is not an obvious requirement.
 
Unitary actions and measurements are the elementary operations allowed on closed quantum systems. However, no quantum system (excepting, possibly, the entire universe) is truly closed. Open quantum systems may be treated as subsystems of some larger closed quantum system. The unitary evolution of the entire closed system induces a (generally non-unitary) evolution on the open quantum subsystem. 
Such operations, which include measurements and unitary actions in their span, are referred to as {\em quantum  operations} in the context of quantum information theory~\cite{bib:niechuang}. The theory of open quantum systems has been developed from the point of view of fundamental physics in the work of Feynman and Vernon, Caldeira and Leggett and Prokof'ev and Stamp~\cite{bib:bath1,bib:bath2,bib:bath3,bib:bath4}. In this seminal work the environment degrees of freedom are included explicitly, and renormalization group arguments are adduced to model the environment as belonging to  one of two universality classes: either the environment degrees of freedom are localized (spin bath) or delocalized (oscillator bath).  ``Random level'' models of the environment have also been studied~\cite{bib:randombath1,bib:randombath2}.
 
In keeping with the motivation of this issue we shall leave the fascinating computational properties of both discrete-time and continuous-time quantum random walks aside. We shall focus instead on the physical interpretation of the classical stochastic lattice gas as a microscopic model for diffusion and of the quantum lattice gas as a microscopic model for a single particle obeying the Dirac equation in the continuum limit.  Our aim in the present paper will be the construction of a single model capable of capturing both types of behavior in different parameter regimes.

We shall consider decoherence arising from an interaction with an environment coupling only to the particles' internal degrees of freedom and not to the particles' position degrees of freedom. Such decoherence models are referred to as ``coin'' decoherence in the context of quantum random walks. In general, coin decoherence produces a transition from behavior characteristic of a quantum walk in which the standard deviation of the walk grows linearly in time to behavior characteristic of a classical walk in which the standard deviation varies as the square root of time~\cite{bib:alagicrussellqrw,bib:kendonqrw1,bib:kendonqrw2,bib:kendonqrw3,bib:alagicrussellqrw,bib:brun1,bib:brun2}. Previous decoherence models for quantum walks considered a scattering step which is a unitary action with probability $1-p$, and a unitary action followed by a projective measurement with probability $p$~\cite{bib:alagicrussellqrw,bib:kendonqrw1,bib:kendonqrw2,bib:kendonqrw3}. Models in which measurements on the coin yield less than total information have also been considered~\cite{bib:kendonsandersqrw}.

We begin by considering a decoherence model for quantum lattice gases (or equivalently, discrete-time quantum random walks) with an explicitly physical motivation. We require that the system-environment interaction be fixed and that the interaction preserve the parity invariance of the original lattice gas. We first describe the unitary lattice-gas dynamics and the extension of such dynamics to the density matrix formalism, necessary for the introduction of quantum operations into such dynamics. We then introduce the framework of quantum operations, and discuss two types of decoherence models for quantum lattice gases. We derive the constraints on the unitary operator coupling the system and environment arising from the requirement of parity invariance. Numerical results are presented for one parameterization of the decoherence model, which qualitatively verify the transition from quantum to classical ({\em i.e.}, diffusive) behavior in the model. We close the paper with discussion and some directions for future work.

\section{Decoherence}

The state vector formulation of quantum mechanics is inadequate to describe situations in which we have imperfect knowledge of the quantum state. Such perfect knowledge is expressed by the system being in a {\em pure state}, that is, a vector of complex amplitudes whose moduli squared sum to one.  Consider the imperfect preparation of a quantum state such that state $\ket{\psi}$ is prepared with probability $P(\ket{\psi})$, where the (classical) probabilities $P(\ket{\psi})$ sum to one. We cannot simply represent this by a real linear convex combination of states, as we already have a complex superposition over observable basis states. We must instead consider a linear real convex  combination of the dyads $\ket{\psi}\bra{\psi}$,
\begin{equation}
\rho=\sum P(\psi) \ket{\psi}\bra{\psi}.
\end{equation}
Such a real linear convex combination is called a {\em mixed state}. The density matrix is the appropriate tool for simultaneously describing two probabilistic aspects of the theory -- one arising from classical uncertainty about the state of the system, and one arising from the fundamental uncertainty arising from quantum superposition.

The time evolution of the density matrix may be obtained by linear extension of the evolution of the pure states. The time evolution of a pure state is given by $\ket{\psi'}=U\ket{\psi}$ where $U$ is the unitary evolution operator of the system, and so the time evolution of the density matrix is given by conjugation: $U$: $\rho'=U\rho U^\dagger$, where dagger indicates the Hermitian conjugate. The Hilbert space of a quantum system which can be divided into two subsystems $A$ and $B$ possesses a basis which can be tensor factored such that each basis vector $\ket{m}$ can be expressed as a tensor product $\ket{m_a}\otimes\ket{m_b}$, such that $\ket{m_a}\in H_a$ and $\ket{m_b}\in H_b$. The reduced density matrix of subsystem $A$ is obtained by taking the partial trace of the full density matrix over subsystem $B$.

We may now define a quantum operation on the density matrix of system $A$. We take the tensor product of the density matrix of the system $A$ with that of the environment $B$.  A unitary operation $U^{AB}$ acts on the resulting density matrix by conjugation. The environmental subsystem $B$ is then traced over, resulting in a new density matrix for the system $A$. Such quantum operations are therefore maps from density matrices to density matrices. Such maps may be constructed without reference to an environment state by invoking the operator-sum representation. The theory of such maps may also be formulated axiomatically without reference to the constructive procedure adduced here~\cite{bib:niechuang}. In the present paper we utilize the unitary representation of quantum operations given above, while noting that a formulation of the noise model we shall construct in terms of the operator-sum representation is possible, and in fact may be a more convenient representation of the model.

\section{Quantum Lattice-Gas Model}

In the present paper we restrict attention to one dimensional lattice gases with two directions per site. The (classical) particle states are specified as follows:  Each site on the lattice has two lattice vectors connecting it to its left and right neighbors. There may be at most one particle per site per vector~\footnote{This is sometimes called the ``exclusion principle'' for lattice gases, but it should be noted that it is unrelated to the Pauli exclusion principle.}. The dynamics of all lattice gases take place in two substeps. First, the particles propagate along their vectors to neighboring sites, retaining their velocity as they do so.  Second, the particles at each site undergo a collision changing the occupations of the vectors at each site.  The propagation step clearly conserves any quantity that is obtained by summing a function of particle mass and velocity over all particles, since those quantities are not changed as particles propagate.  The collision step is required to conserve a subset of these quantities that are of physical interest, such as mass, momentum, etc.  In the following we shall consider models which conserve particle number only; if we regard the particles as each having unit mass, this may be thought of as conservation of mass.

For a one dimensional lattice with two vectors per site, the only deterministic rules which preserve particle number are trivial. The first nontrivial model occurs when one considers a stochastic collision in which a single particle at a site has probability $1-p$ to reverse direction. The stochastic lattice gas with a single particle may be identified with a random walk. Generalizing to multiple particles we find the evolution of the single particle distribution function for a classical stochastic lattice gas obeys the diffusion equation. The stochastic models include the (trivial) deterministic models as the special case $p=1$.

The single timestep evolution operator  $U$  of the quantum lattice gas without decoherence is the composition of advection and scattering steps:
\begin{equation}
\begin{split}
\sum\psi_{x,\alpha} |x,\alpha\rangle\underrightarrow{\rm advect}& \sum\psi_{x,\alpha}|x+\alpha,\alpha\rangle\\
\underrightarrow{\rm scatter}
 & \sum\psi_{x,\alpha} S_{\alpha\alpha'} |x+\alpha,\alpha'\rangle,  \cr
\end{split}
\end{equation}
where the scattering matrix may be parameterized up to a global phase as: 
\begin{equation}\label{eq:meyercoll}
S = \begin{pmatrix}\cos\theta & i\sin\theta \\
             i\sin\theta &  \cos\theta \end{pmatrix}
\end{equation}
and the Hilbert space is $2N$ dimensional, where $N$ is the lattice size. This quantum lattice gas was shown to yield the continuum propagator for the Dirac equation for a particle with mass $\tan \theta$ in~\cite{bib:meyer2}. The model may also be interpreted as a discrete-space discrete-time quantum random walk, subject to the constraint that the scattering rule be parity invariant.

We generalize the above model by first extending the dynamics to those of the density matrix for the lattice gas. This allows us to handle the mixed states which will arise when we introduce decoherence into the dynamics. We couple the particles' internal degrees of freedom to a bath of arbitrary size and act on the internal degrees of freedom and the bath with a unitary matrix which is the product:
\begin{equation}
U_{c}\cdot \left[S\otimes{\mathbb{I}}_{bath}\right],
\end{equation}
where $\mathbb{I}$ is the identity operator on the bath. We then trace over the environment degrees of freedom at each timestep.

This decoherence model corresponds to a generalization of the quantum lattice gas to the case where the collision operator is neither a unitary operator, nor a classical Markov matrix as in the stochastic classical lattice gas, but a quantum operation. The set of quantum operations contains both unitary actions and Markov actions as special cases. The unitary operator is clearly included as a special case when one considers $U_{c}$ which do not couple the system and the environment ({\em i.e.}, which tensor factor into $U_c=U_{c(sys)}\otimes U_{c(bath)}$). 

It is clear that the Markov operations of the classical stochastic lattice gas are included when one considers an initialization of the density matrix of the lattice gas in a completely classical state -- that is, a density matrix at each site with only diagonal entries, $p_l$ and $p_r$.  The action of the unitary part of the collision operator on this matrix is:
\begin{equation}
 \begin{pmatrix}\cos\theta & i\sin\theta \\
             i\sin\theta &  \cos\theta \end{pmatrix} \begin{pmatrix}p_r & 0 \\
            0 & p_l\end{pmatrix} \begin{pmatrix}\cos\theta & -i\sin\theta \\
             -i\sin\theta &  \cos\theta \end{pmatrix}
             = \begin{pmatrix}p_l\cos^2\theta + p_r\sin^2\theta  & i\sin\theta\cos\theta(p_r-p_l) \\
            i\sin\theta\cos\theta(p_l-p_r) & p_l\sin^2\theta + p_r\cos^2\theta \end{pmatrix}
\end{equation}
This reproduces the action of a classical Markov matrix on  the vector of probabilities $(p_l,p_r)$, where the probability that the particle continues in its current state is $\cos^2\theta$ if the interaction with the environment induces the map:
\begin{equation}\label{eq:measure}
 \begin{pmatrix}p_l\cos^2\theta + p_r\sin^2\theta  & i\sin\theta\cos\theta(p_r-p_l) \\
            i\sin\theta\cos\theta(p_l-p_r) & p_l\sin^2\theta + p_r\cos^2\theta \end{pmatrix}
\mapsto \begin{pmatrix}p_l\cos^2\theta + p_r\sin^2\theta  &0 \\
            0 & p_l\sin^2\theta + p_r\cos^2\theta \end{pmatrix}.
\end{equation}

We can construct this mapping by coupling the system to a two-dimensional environment in the completely mixed state $1/2\mathbb{I}$ using the controlled-{\tt NOT} gate where the system is the control qubit. More generally, the map given in Eq.~(\ref{eq:measure}) is the effect of a measurement on the system, and such actions are included in the set of quantum operations. The correspondence of unitary operations combined with measurements as equivalent to Markov operations is also discussed in the context of Type-II quantum computing in~\cite{bib:lovetII}.

In order to complete the demonstration that classical stochastic lattice gases are included as a subset of models described by a density matrix whose collision process is a quantum operation, we must show that a classical ({\em i.e.}, diagonal) density matrix evolves to another classical density matrix under propagation.  The classical density matrix is non-zero only in entries $\ket{x,\alpha}\bra{x,\alpha}$, which evolve under propagation to $\ket{x+\alpha,\alpha}\bra{x+\alpha,\alpha}$.  Diagonal density matrices evolve to diagonal density matrices under propagation and a subset of collision operations evolving diagonal density matrices to diagonal density matrices are equivalent to the Markov matices implementing collisions of a classical stochastic lattice gas.

Hence a single-particle quantum lattice-gas simulation in which the entire density matrix is stored and in which the collision rule is a quantum operation includes as special cases quantum lattice-gas evolution and the stochastic evolution of a classical lattice gas. Deterministic classical lattice gases are included as they are a special case of stochastic lattice gases.  Such a model therefore provides a bridge between the quantum lattice gas which possesses the Dirac equation as a continuum limit  and classical stochastic lattice gases which possess the diffusion equation as a continuum limit. 

\section{Parity-Preserving Noise}

We now define the set of quantum operations giving our new scattering rule. Specification of a set of quantum operations defines a ``quantum noise'' model. Such a model has two ingredients: a model for the environment state and a model for the system-environment interaction. Before specifying the environment state we must specify how many dimensions the environment Hilbert space must have. Here we may invoke a theorem which states that for a $d$-dimensional system Hilbert space a $d^2$-dimensional environment is sufficient to produce every possible quantum operation on the system. This theorem states that for a quantum operation specified by a set of principal components in the operator-sum representation it is always possible to find a unitary operator coupling a $d^2$-dimensional environment to a $d$-dimensional system which reproduces this quantum operation~\cite{bib:niechuang}. 

This suggests the following noise model: We initialize the four-dimensional environment in a fiducial state ($|00\rangle$ for example). We then sample from a distribution over the unitary group $U(8)$ and apply our sampled operator to the eight-dimensional system-environment pair, and then we trace over the environment. Such a model samples from the entire set of quantum operations. The preparation of the environment in a fiducial state does not imply a loss of generality, as the sampled operation can be considered to be first an operator acting only on the environment preparing a random environment state, followed by an operation coupling the system and environment. The distribution over the set of quantum  operations is induced in a nontrivial way by the distribution over $U(8)$. Such a model would resemble a random level, or random matrix model, of the environment~\cite{bib:randombath1,bib:randombath2}.   

On the other hand, our aim is the construction of a noise model with physically inspired constraints on the system-environment interaction.  For such physically motivated noise models we wish the system-environment coupling matrix to be a constant unitary operator, arising from a putative fixed system-environment interaction Hamiltonian (which we may or may not know).  If we regard our quantum lattice gas as a discrete model for the Dirac equation, we note that the interaction of the helicity degree of freedom of a Dirac particle with an environment is indeed fixed by fundamental physics.

We must take care about the meaning of the theorem invoked above for noise models with a constant system-environment interaction. The theorem does not state that a fixed unitary operator coupling a $d^2$-dimensional environment to a $d$-dimensional system can reproduce every quantum operation on the system. In the sequel we construct our noise model for an environment of arbitrary dimension, although we revert to a four-dimensional environment for reasons of computational tractability for simulations.

Equation~(\ref{eq:meyercoll}) gives an explicit parameterization of all parity-preserving two-dimensional unitary operators, up to a global phase. Such convenient parameterizations of quantum operations do not yet exist, and so we follow the explicit constructive procedure for such operations given above.
We choose a model for the environment such that its state is a unimodular complex vector whose Cartesian components are independent random variables. The environment-system interaction is fixed, and we consider a wide class of such interactions, namely those which preserve the parity invariance of the original unitary quantum lattice gas. The unitary update $S$ obeys parity invariance $ST=TS$ where $T$ is the parity exchange operator. The parity preserving couplings $U_s$ have the property that they commute with the parity exchange operator acting on the system tensored with the identity operator acting on the environment. 
\begin{equation}\label{eq:timeevolsym}
[U_s,T\otimes\mathbb{I}]=0
\end{equation}

Eq.~(\ref{eq:timeevolsym}) expresses a discrete symmetry of a discrete-time evolution operator. In physics we are more usually concerned with continuous symmetries and continuous time evolution operators. The usual statement of invariance of an interaction Hamiltonian under a particular symmetry transformation is that the infinitesimal generators of the symmetry transformation commute with the Hamiltonian. In the language of Lie groups this means that the Hamiltonian lies in the commutator subalgebra of the generators of the symmetry transformation in the Lie algebra of $U(N)$, where $N$ is the number of degrees of freedom of our system. We may also apply these ideas to a discrete symmetry of a discrete-time discrete-space model. Let $t$ be the Lie algebra element corresponding to $T\otimes\mathbb{I}$.  Let $u$ be the Lie algebra element corresponding to $U$.  A sufficient condition that Eq.~(\ref{eq:timeevolsym}) holds is:
\begin{equation}
[u,t]=0.
\end{equation}

The Lie algebra of $U(N)$ is the set of anti-hermitian matrices.  We use the basis arising from the root system of the Lie algebra of $U(N)$~\cite{bib:brockertomdieck}:
\begin{equation}
\begin{split}
D^p _{kl}&= i\delta_{pk}\delta_{pl}\hspace{1cm} 1\leq p\leq N\\
S^{qp}_{kl}&=i(\delta_{kp}\delta_{ql}+\delta_{kq}\delta_{pl})  \hspace{1cm} 1\leq p\leq N\hspace{1cm} q<p \\
A^{qp}_{kl}&=(\delta_{kp}\delta_{ql}-\delta_{kq}\delta_{pl})  \hspace{1cm} 1\leq p\leq N\hspace{1cm} q<p \\
\end{split}
\end{equation}\label{eq:lab}
where $\delta_{xy}$ is the Kronecker delta, and there is no sum on repeated indices. We note that $A^{qq}=0$ and $S^{qq}=2D^q$. The convention for the antisymmetric matrices is chosen so that the labelling superscripts increase from left to right, and so that the negative entry is always in the upper triangular portion of the matrix. 

The commutation relations of the Lie algebra follow from~\ref{eq:lab}:
\begin{equation}
\begin{split}
[D^p,D^q] _{kl}&=0\hspace{1cm} 1\leq p\leq N\\
[D^r,S^{qp}]_{kl}&=\delta_{pr}A^{qr}+\delta_{rq}A^{pr}\\
[D^r,A^{qp}]_{kl}&=\delta_{rp}S^{rq}-\delta_{rq}S^{rp}\\
[S^{rs},A^{qp}]_{kl}&=\delta_{sp}S^{rq}-\delta_{sq}S^{rp}+\delta_{rp}S^{sq}-\delta_{rq}S^{sp}\\
[S^{rs},S^{qp}]_{kl}&=\delta_{ps}A^{qr}+\delta_{qs}A^{pr}+\delta_{pr}A^{qs}+\delta_{qr}A^{ps}\\
[A^{rs},A^{qp}]_{kl}&=\delta_{ps}A^{qr}+\delta_{qs}A^{pr}+\delta_{pr}A^{qs}+\delta_{qr}A^{ps}
\end{split}
\end{equation}

The block diagonal form of $T\otimes\mathbb{I}$ makes it straightforward to diagonalize, and it is therefore straightforward to obtain the Lie algebra element $t$.
\begin{equation}
t=\ln T\otimes 1=\frac{\pi}{2}\sum_{r=1}^N D^r - \frac{\pi}{2}\sum_{s=1}^{N/2}S^{(2s-1)2s} 
\end{equation}
A general element $u$ of the Lie algebra may be written:
\begin{equation}
u=\sum_{r=1}^{N}\alpha_rD^r + \frac{1}{2}\sum_{p=1}^N\sum_{q=1}^N \left[\beta_{qp} S^{qp} + \gamma_{qp} A^{qp}\right]
\end{equation}
where $\beta_{qp}=\beta_{pq}$ and $\gamma_{qp}=-\gamma_{pq}$, and $\beta_{qq}=\gamma_{qq}=0$. 

The commutator is then
\begin{equation}
\begin{split}
[u,t]&=\frac{\pi}{2}\left[\sum_{s=1}^N D^s- \sum_{s=1}^{N/2}S^{(2s-1)2s}  ,\sum_{r=1}^{N}\alpha_rD^r +\frac{1}{2}\sum_{p=1}^N\sum_{q=1}^N \left[\beta_{qp} S^{qp} + \gamma_{qp} A^{qp}\right]\right]\\
\end{split}
\end{equation}
Applying the structure constants of the Lie algebra and using $\beta_{qp}=\beta_{pq}$ and $\gamma_{qp}=-\gamma_{pq}$ gives:
\begin{equation}
\begin{split}
[u,t]&=-\frac{\pi}{2}\sum_{s=1}^{N/2}\sum_{p=1}^N\biggl[\beta_{p(2s)}A^{p(2s-1)}+\beta_{p(2s-1)}A^{p(2s)}+\gamma_{p(2s)}S^{(2s-1)p}+\gamma_{p(2s-1)}S^{(2s)p} \biggr]\\
\end{split}
\end{equation}
The constraint that this be zero imposes a set of constraints on the $\beta$ coefficients and a set of constraints on the $\gamma$ coefficients. Because the coefficients and the matrices $A$ are real, while the matrices  $S$ are pure imaginary, we may rearrange terms involving the $A$'s and  $S$'s separately to obtain these constraints.  We write
\begin{equation}
[u,t] = C_\beta + C_\gamma.
\end{equation} 
Where $C_\beta=0$ and $C_\gamma=0$ are necessary conditions for $[u,t]=0$. 
\begin{equation}\label{eq:cbeta}
\begin{split}
C_\beta&=-\frac{\pi}{2}\sum_{s=1}^{N/2}\sum_{p=1}^N\biggl[\beta_{p(2s)}A^{p(2s-1)}+\beta_{p(2s-1)}A^{p(2s)}\biggr]\\
&=-\frac{\pi}{2}\sum_{s\phantom{=}even}^{N}\sum_{p\phantom{=}even}^N\beta_{ps}A^{p(s-1)}-\frac{\pi}{2}\sum_{s\phantom{=}even}^{N}\sum_{p\phantom{=}odd}^{N-1}\beta_{ps}A^{p(s-1)}\\
&-\frac{\pi}{2}\sum_{s\phantom{=}odd}^{N-1}\sum_{p\phantom{=}even}^N\beta_{ps}A^{p(s+1)}-\frac{\pi}{2}\sum_{s\phantom{=}odd}^{N-1}\sum_{p\phantom{=}odd}^{N-1}\beta_{ps}A^{p(s+1)}\\
\end{split}
\end{equation}
We wish to rearrange terms so that we have a unique set of $A$'s whose coefficients we can set to zero in order to obtain our constraints.  In the first and fourth terms here the indices on the $A$'s have opposite parity, whereas in the second and third terms the indices have the same parity in each term, but the indices are both odd in the second term and both even in the third term. This means that the first and fourth terms may be combined by exchanging dummy indices and using the symmetry properties of the $A$'s, whereas the second and third terms must be dealt with separately.  Denoting term $x$ in the right hand side of Eq.~(\ref{eq:cbeta}), $C_\beta^x$, and taking the first and fourth terms:
\begin{equation}
\begin{split}
C^1_\beta+C^4_\beta=-\frac{\pi}{2}\sum_{s\phantom{=}odd}^{N-1}\sum_{p\phantom{=}even}^N\beta_{p(s+1)}A^{ps}-\frac{\pi}{2}\sum_{s\phantom{=}even}^{N}\sum_{p\phantom{=}odd}^{N-1}\beta_{p(s-1)}A^{ps}\\
\end{split}
\end{equation}
Exchanging $p$ and $s$ in the second term on the right hand side:
\begin{equation}
\begin{split}
C^1_\beta+C^4_\beta=-\frac{\pi}{2}\sum_{s\phantom{=}odd}^{N-1}\sum_{p\phantom{=}even}^N\left[\beta_{p(s+1)}A^{ps}+\beta_{s(p-1)}A^{sp}\right]\\
\end{split}
\end{equation}
Using the antisymmetry of the $A$'s, we have
\begin{equation}
\begin{split}
C^1_\beta+C^4_\beta=-\frac{\pi}{2}\sum_{s\phantom{=}odd}^{N-1}\sum_{p\phantom{=}even}^N\left[\beta_{p(s+1)}-\beta_{s(p-1)}\right]A^{ps}\\
\end{split}
\end{equation}
Now consider the second and third terms in $C_\beta$. The second term is:
\begin{equation}
\begin{split}
C_\beta^2&=-\frac{\pi}{2}\sum_{s\phantom{=}odd}^{N-1}\sum_{p\phantom{=}odd}^{N-1}\beta_{p(s+1)}A^{ps}\\
&=-\frac{\pi}{4}\sum_{s\phantom{=}odd}^{N-1}\sum_{p\phantom{=}odd}^{N-1}\left[\beta_{p(s+1)}-\beta_{s(p+1)}\right]A^{ps}\\
\end{split}
\end{equation}
The third term is:
\begin{equation}
\begin{split}
C_\beta^3&=-\frac{\pi}{2}\sum_{s\phantom{=}even}^{N}\sum_{p\phantom{=}even}^N\beta_{p(s-1)}A^{ps}\\
&=-\frac{\pi}{4}\sum_{s\phantom{=}even}^{N}\sum_{p\phantom{=}even}^N\left[\beta_{p(s-1)}-\beta_{s(p-1)}\right]A^{ps}\\
\end{split}
\end{equation}
These equations give the following set of conditions on the beta coefficients:
\begin{equation}
\begin{split}
\beta_{p(s+1)}-\beta_{s(p-1)}=0\hspace{.5cm}s\mod2=1\hspace{1cm}p\mod2=0\\
\beta_{p(s+1)}-\beta_{s(p+1)}=0\hspace{.5cm}s\mod2=1\hspace{1cm}p\mod2=1\\
\beta_{p(s-1)}-\beta_{s(p-1)}=0\hspace{.5cm}s\mod2=0\hspace{1cm}p\mod2=0\\
\end{split}
\end{equation}
These constraints are redundant. In fact a pair of constraints is sufficient to ensure $C_\beta=0$:
\begin{equation}
\begin{split}
&\beta_{p(s+1)}-\beta_{s(p-1)}=0\hspace{.5cm}s\mod2=1\hspace{1cm}s< p\\
&\beta_{p(s+1)}-\beta_{s(p+1)}=0\hspace{.5cm}s\mod2=1\hspace{1cm}p\mod2=1\hspace{1cm}s< p
\end{split}
\end{equation}
We now consider $C_\gamma$.
\begin{equation}
\begin{split}
C_\gamma&=-\frac{\pi}{2}\sum_{s=1}^{N/2}\sum_{p=1}^N\biggl[\gamma_{p(2s)}S^{(2s-1)p}+\gamma_{p(2s-1)}S^{(2s)p} \biggr]\\
&=-\frac{\pi}{2}\sum_{s\phantom{=}even}^{N}\sum_{p\phantom{=}even}^N\gamma_{p(s-1)}S^{sp}-\frac{\pi}{2}\sum_{s\phantom{=}even}^{N}\sum_{p\phantom{=}odd}^{N-1}\gamma_{p(s-1)}S^{sp}\\
&-\frac{\pi}{2}\sum_{s\phantom{=}odd}^{N-1}\sum_{p\phantom{=}even}^N\gamma_{p(s+1)}S^{sp}-\frac{\pi}{2}\sum_{s\phantom{=}odd}^{N-1}\sum_{p\phantom{=}odd}^{N-1}\gamma_{p(s+1)}S^{sp}\\
\end{split}
\end{equation}
Exchanging dummy indices in the third term, expanding the first and fourth terms, and utilizing the symmetry of the $S$'s and the antisymmetry of the $\gamma$'s gives:
\begin{equation}
\begin{split}
C_\gamma&=-\frac{\pi}{4}\sum_{s\phantom{=}even}^{N}\sum_{p\phantom{=}even}^N\left[\gamma_{p(s-1)}+\gamma_{s(p-1)}\right]S^{sp}\\
&-\frac{\pi}{2}\sum_{s\phantom{=}even}^{N}\sum_{s\phantom{=}odd}^{N-1}\left[\gamma_{p(s-1)}-\gamma_{(p+1)s}\right]S^{sp}\\
&-\frac{\pi}{4}\sum_{s\phantom{=}odd}^{N-1}\sum_{p\phantom{=}odd}^{N-1}\left[\gamma_{p(s+1)}+\gamma_{s(p+1)}\right]S^{sp}\\
\end{split}
\end{equation}
This yields the constraints on the $\gamma$'s:
\begin{equation}
\begin{split}
&\gamma_{p(s-1)}-\gamma_{(p+1)s}=0\hspace{.5cm}s\mod2=0\hspace{1cm}p\mod2=1\hspace{1cm}p<s-1\\
&\gamma_{p(s-1)}+\gamma_{s(p-1)}=0\hspace{.5cm}s\mod2=0\hspace{1cm}p\mod2=0\hspace{1cm}p\leq s\\
\end{split}
\end{equation}

\section{Simulations}

Simulations were performed for a system whose internal degree of freedom is coupled to a four-dimensional bath.  The coupling operator is therefore an element of $U(8)$. The general form of an element of the Lie algebra of $U(8)$ obeying the constraints derived above is
\begin{equation}\label{eq:sym}
\begin{split}
i\begin{pmatrix}
\alpha_1&\beta_{12}&\beta_{13}&\beta_{14}&\beta_{15}&\beta_{16}&\beta_{17}&\beta_{18}\\
\beta_{12}&\alpha_2&\beta_{14}&\beta_{13}&\beta_{16}&\beta_{15}&\beta_{18}&\beta_{17}\\
\beta_{13}&\beta_{14}&\alpha_3&\beta_{34}&\beta_{35}&\beta_{36}&\beta_{37}&\beta_{38}\\
\beta_{14}&\beta_{13}&\beta_{34}&\alpha_4&\beta_{36}&\beta_{35}&\beta_{38}&\beta_{37}\\
\beta_{15}&\beta_{16}&\beta_{35}&\beta_{36}&\alpha_5&\beta_{56}&\beta_{57}&\beta_{58}\\
\beta_{16}&\beta_{15}&\beta_{36}&\beta_{35}&\beta_{56}&\alpha_6&\beta_{67}&\beta_{57}\\
\beta_{17}&\beta_{18}&\beta_{37}&\beta_{38}&\beta_{57}&\beta_{67}&\alpha_7&\beta_{78}\\
\beta_{18}&\beta_{17}&\beta_{38}&\beta_{37}&\beta_{58}&\beta_{57}&\beta_{78}&\alpha_8\\
\end{pmatrix}
+\begin{pmatrix}
0&0&-\gamma_{13}&-\gamma_{14}&-\gamma_{15}&-\gamma_{16}&-\gamma_{17}&-\gamma_{18}\\
0&0&-\gamma_{14}&-\gamma_{13}&\gamma_{16}&-\gamma_{15}&-\gamma_{18}&-\gamma_{17}\\
\gamma_{13}&\gamma_{14}&0&0&-\gamma_{35}&-\gamma_{36}&-\gamma_{37}&-\gamma_{38}\\
\gamma_{14}&\gamma_{13}&0&0&-\gamma_{36}&-\gamma_{35}&-\gamma_{38}&-\gamma_{37}\\
\gamma_{15}&\gamma_{16}&\gamma_{35}&\gamma_{36}&0&0&-\gamma_{57}&-\gamma_{58}\\
\gamma_{16}&\gamma_{15}&\gamma_{36}&\gamma_{35}&0&0&-\gamma_{58}&-\gamma_{57}\\
\gamma_{17}&\gamma_{18}&\gamma_{37}&\gamma_{38}&\gamma_{57}&\gamma_{58}&0&0\\
\gamma_{18}&\gamma_{17}&\gamma_{38}&\gamma_{37}&\gamma_{58}&\gamma_{57}&0&0\\
\end{pmatrix}
\end{split}
\end{equation}

In all cases the system was initialized in a pure state corresponding to a gaussian spatial wavefunction centred at the origin with a standard deviation equal to one quarter of the lattice size, with equal amplitudes for both internal states of the particle. The system was a periodic lattice with $64$ sites, and the parameter $\theta$ in the unitary part of the collision operator was set equal to $0.35$.

Three types of simulations were performed.  First, a simulation of the density matrix of the system and bath was performed in which the coupling matrix was the identity operator. In this case the quantum lattice gas reproduces the unitary evolution expected in the case that the system-bath coupling is zero. Second, simulations were performed in which the system-bath coupling operator is the Fourier transform over the cyclic group $\mathbb{Z}_8$. Finally, simulations were performed in which all coefficients $\alpha$, $\beta$ and $\gamma$ in a matrix of the form~(\ref{eq:sym}) were set equal to one and the resulting matrix numerically exponentiated to obtain a parity-preserving coupling.  

The simulation with no system-bath coupling shows typical unitary evolution on a cycle, with the wave-packet dispersing until the edges of the wave-packet reach the periodic boundaries of the system, at which time interference occurs between the original packet and the reentrant components. The reversibility and unitarity of the dynamics is apparent as the system never settles into a static equilibrium state. The simulations in which the system-bath coupling is given by the Fourier transform, which violates parity invariance, exhibit a driving of the system to the right.  Additional simulations in which the coupling is given by the conjugate of the Fourier transform with the parity inversion operator show the same driving effect in the opposite direction, as expected. Simulations in which the system is coupled to the bath by our example parity-invariant unitary matrix show no driving effect. The system undergoes an irreversible evolution of the initial probability distribution to the uniform distribution on the cycle.

The behavior of the model shows its intermediate quantum-classical nature in two ways.  First, the decay of the probability distribution shows some residual ``wave-like'' behavior in addition to the overall damping. Second, the final density matrix is not a completely mixed state, but retains non-zero off-diagonal components within a finite band. Both these effects indicate that the coupling matrix chosen implements a coupling strong enough to cause irreversible dynamics on an observable timescale but weak enough that the dynamics retains some interesting quantum characteristics. 

\begin{figure}[htp]
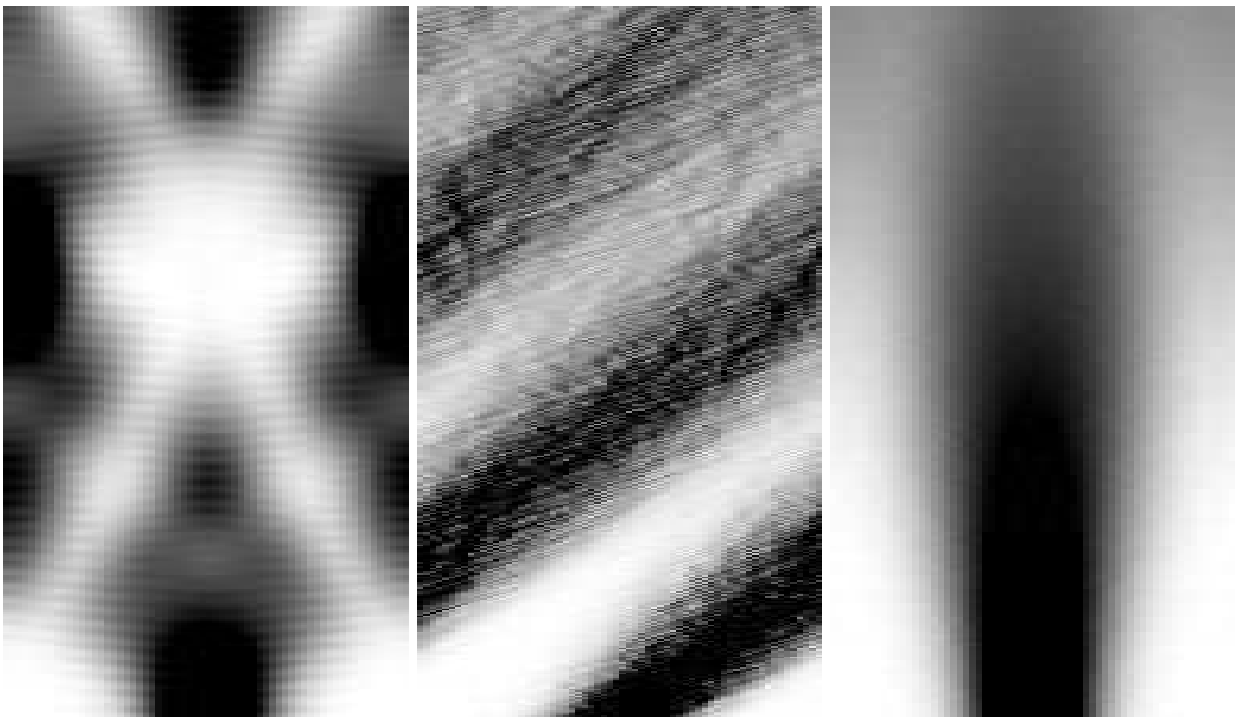

\centering
\includegraphics[width=0.3\textwidth,height=0.4\textheight]{figure1a.epsf}
\includegraphics[width=0.3\textwidth,height=0.4\textheight]{figure1b.epsf}
\includegraphics[width=0.3\textwidth,height=0.4\textheight]{figure1c.epsf}
\caption[]{Plots of the time evolution of the diagonal components for 400 time steps . The greyscale indicates the magnitude of the density matrix evolving from a Gaussian pure state centered at the origin with standard deviation one quarter of the system size. Left: Unitary evolution. Middle: Evolution with $\mathbb{Z}_8$ coupling the system to the bath. Right: Evolution with a parity-invariant coupling between system and bath.}
\end{figure}

\begin{figure}[htp]
\centering
\resizebox{8cm}{!}{\includegraphics[angle=-90]{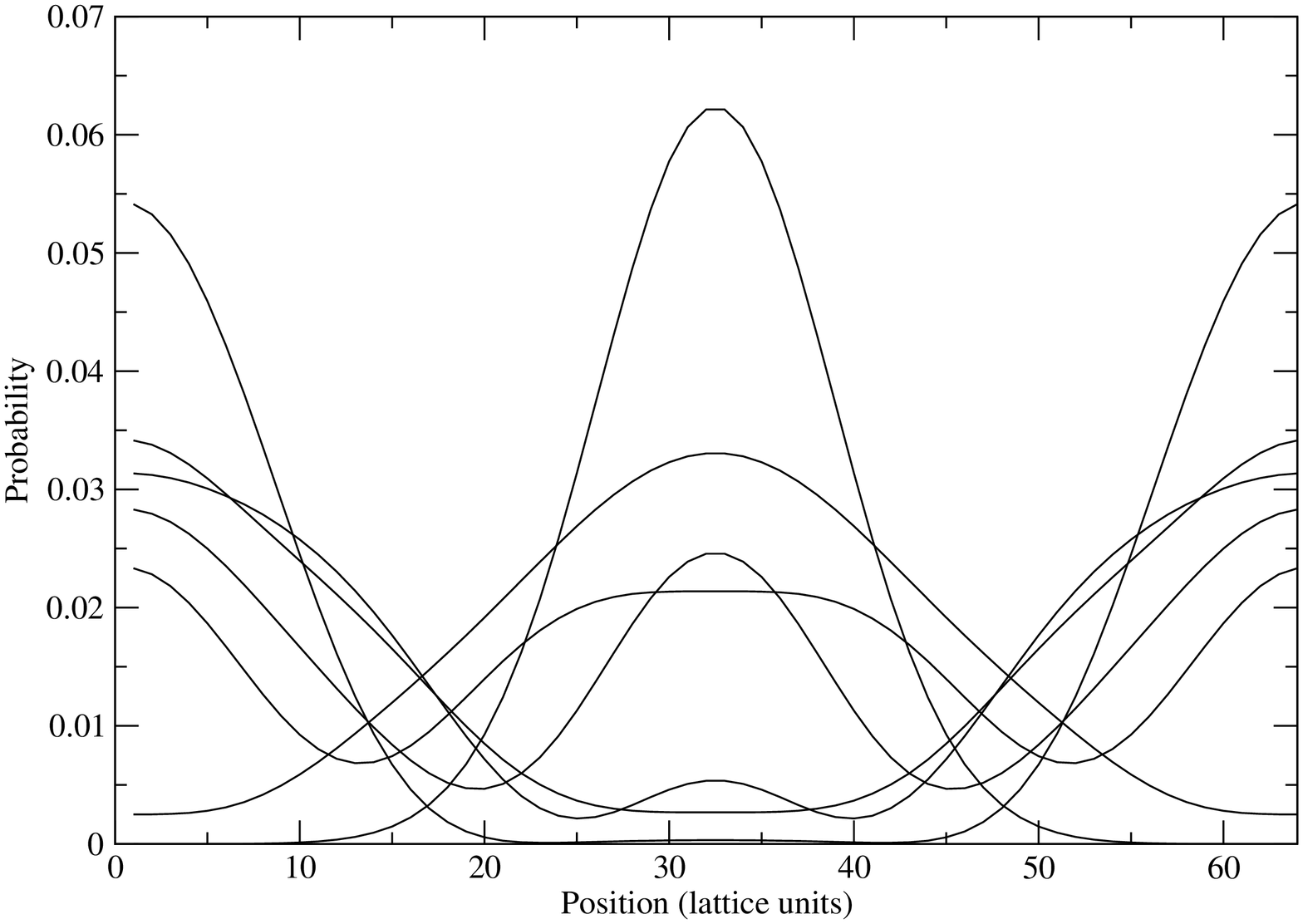}}
\resizebox{8cm}{!}{\includegraphics[angle=-90]{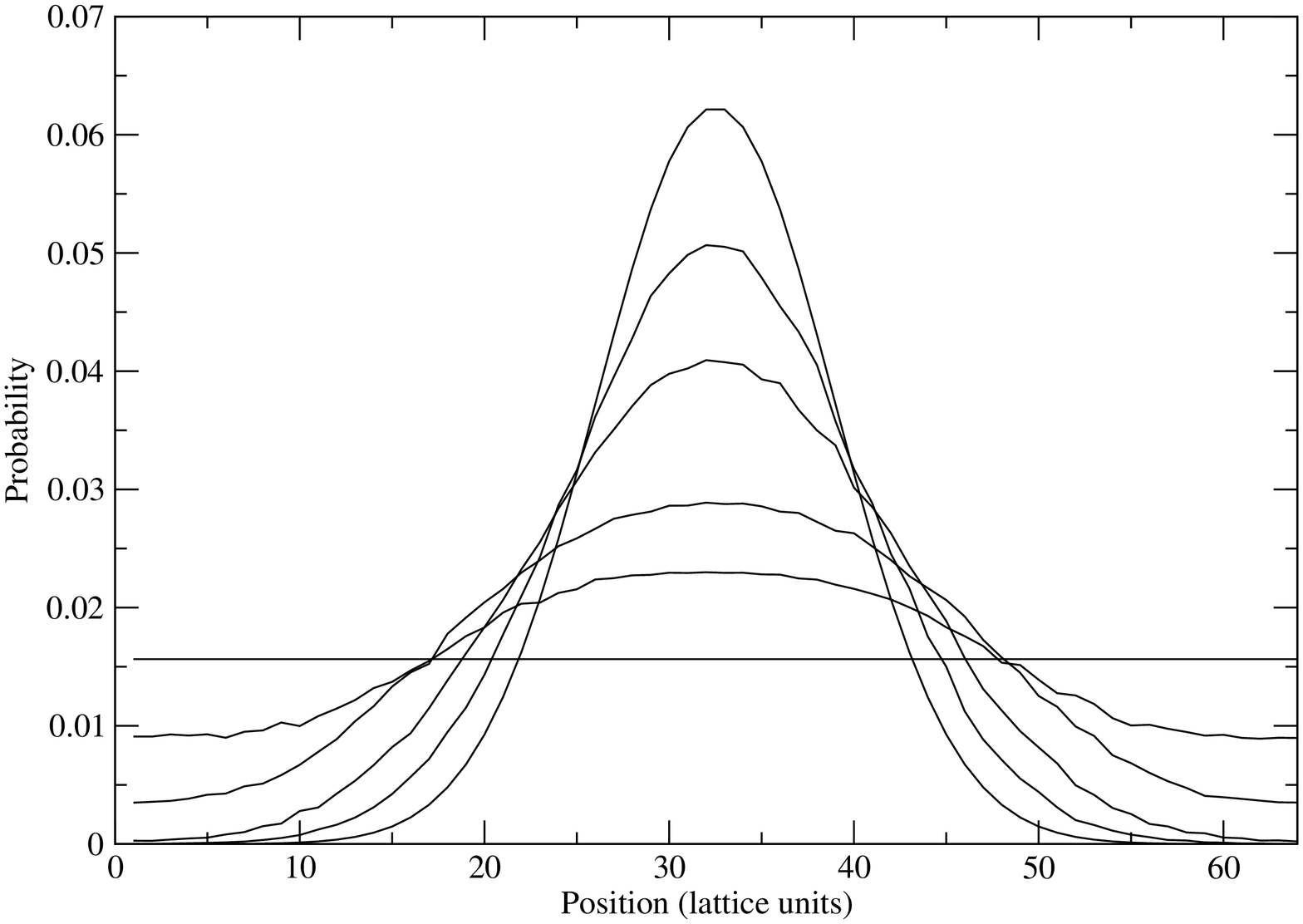}}
\caption[]{Left: Time evolution of the probability distribution for the first $300$ timesteps of unitary evolution. Curves are plotted every $50$ timesteps. Right: Time evolution of the first $300$ timesteps of the parity-invariant decoherent evolution and the final uniform distribution. Curves are plotted every $50$ timesteps, and the final state after $8000$ timesteps of evolution is shown.}
\end{figure}

\section{Conclusions}

We have defined a new quantum lattice-gas model for a single particle (or equivalently, a discrete-space discrete-time quantum random walk) in which the scattering rule is given by a quantum operation, rather than a unitary, deterministic, or stochastic operation. We showed that the model so defined includes unitary, stochastic and deterministic models as special cases, as well as interesting intermediate behavior. Preliminary simulation results confirm this by exhibiting non-unitary diffusive decay of an initial gaussian pure state.

The noise model chosen here utilized a fixed unitary operator coupling a four-dimensional bath to the two-dimensional Hilbert space of the internal degree of freedom of a single quantum lattice-gas particle.
This distinguishes our work from the noise model given by Kendon and Sanders~\cite{bib:kendonsandersqrw} in which a single environment qubit is coupled by an interaction with a strength tunable from the case of no coupling to the case where the environment produced a projective measurement of the coin degree of freedom of the quantum random walk. 

As noted above, a two-dimensional environment with a fixed interaction is insufficient to reproduce all quantum operations.  However, the constraints of parity invariance on the coupling operator were obtained here for $U(N)$, and so this work could be extended to include an arbitrarily large environment. If the most general noise model is desired, the sampling procedure discussed but not implemented above, in which randomly sampled unitary operators couple a four-dimensional bath to the internal degree of freedom of the particle, provably includes all quantum operations. The parameterization of the noise model then involves a parameterization of the distribution of quantum operations induced by a given distribution on the unitary group.

The above discussion motivates several directions for future work.  First, the noise model presented here is not conveniently parameterized. Ideally we would be able to smoothly vary the degree of coupling between the system and the environment from zero (where the time evolution would approximate the Dirac equation) to the case where the environment performs projective measurement of the particles' internal degrees of freedom. This property is possessed by the noise model of Kendon and Sanders~\cite{bib:kendonsandersqrw}, and such a parameterization of the model presented here is certainly possible, although it may be tedious in practice. The work of Kendon and Sanders~\cite{bib:kendonsandersqrw} discusses decoherence in quantum random walks from the point of view of complementarity. The work of the present paper was motivated instead by the principle of correspondence. The generalization here satisfies the requirement that the results agree with classical theory in the case that the particle is strongly coupled to an environment.

The two most natural generalizations of the work described here are to quantum lattice-gas models with multiple particles, and to models defined in multiple dimensions. The constraint of parity invariance becomes the constraint of invariance under discrete rotations in multiple dimensions, and one expects the constraints analogous to those derived here to be correspondingly more complex.  Simulations of multiple particles, even in one dimension, have a classical computational cost which grows exponentially with the number of particles. However, one expects that few-particle simulations will be tractable.

The possibility of efficient quantum simulation of classical systems is an important open problem in the field of quantum computation for physical modeling. One of the central problems in this area is that most of the equations of classical physics of practical interest are irreversible macroscopic equations of motion. The work presented here shows that irreversibility may be simply included by the use of quantum operations instead of unitary matrices. The emergence of irreversible behavior in the degrees of freedom of the subsystem exhibited here is a manifestation of the ``arrow of time'' of non-equilibrium thermodynamics. The interesting practical question, which remains open, is whether there are systems for which the time complexity of such classical and quantum simulation is different.  More glibly:  Can time's arrow may be made to move faster on a quantum computer?

\section{Acknowledgements}
PJL and BMB were supported by DARPA QuIST program administered under AFOSR grant number F49620-01-1-0566, and by ARO contract number W911NF-04-1-0334.  BMB was also supported by AFOSR award number FA9550-04-1-0176. Both authors would like to thank AFOSR for their hospitality at the Quantum Computation for Physical Modeling Workshop on Marthas Vineyard in 2004.  The authors have great pleasure in thanking David Meyer, Gianluca Caterina, Howard Brandt, and Seth Lloyd for helpful discussions and questions.


\end{document}